\documentclass[12pt,preprint]{aastex}
\usepackage{lscape}

\def\masyr{{\rm mas}\,{\rm yr}^{-1}}

\def\lim{{\rm lim}}

\begin{document}

\title{New {\it Hipparcos}-based Parallaxes for 424 Dim Stars}

\author{Andrew Gould and Julio Chanam\'e}
\affil{Department of Astronomy, The Ohio State University,
140 W.\ 18th Ave., Columbus, OH 43210}
\authoremail
{gould,jchaname@astronomy.ohio-state.edu}

\singlespace

\begin{abstract}

We present a catalog of 424 common proper motion companions to 
{\it Hipparcos} stars with good ($>3\,\sigma$) parallaxes, thereby
effectively providing new parallaxes for these companions.  Compared
to stars in the {\it Hipparcos} catalog, these stars are substantially
dimmer.  The catalog includes 20 WDs and an additional 29 stars with
$M_V>14$, the great majority of the latter being M dwarfs.

\end{abstract}
 
\section{Introduction
\label{sec:intro}}

{\it Hipparcos} \citep{hip} revolutionized astrometry in three ways.  First, 
its mass-production mode increased the sheer number of stars with mas
parallaxes by several orders of magnitude to $10^5$.  Second, it
directly obtained absolute parallaxes rather than having to convert
from relative to absolute parallax as is necessary from the ground.
Third, it measured homogeneous parallaxes over the whole sky.

However, the {\it Hipparcos} catalog is notably deficient in dim stars,
containing only 3 stars with $M_V\geq 14$ and only 14 with $M_V>13$.  
While the project attempted to 
observe all known dim stars down to its operational limit $V\sim 12$,
at $M_V=14$ this limit corresponds to a distance of 4 pc.  Hence,
dim-star parallaxes generally continue to require the painstaking 
one-at-a-time methodology that had characterized this
subject for the two centuries prior to {\it Hipparcos}.  The classic work
of \citet{monet92} remains crucial for tracing out the bottom
of the main sequence (MS) and the cool subdwarfs (SDs).  \citet{bergeron01}
assembled a catalog of about 150 white dwarfs (WDs) with parallaxes, the
vast majority from the ground.  \citet{jao} have targeted almost 200
nearby stars for parallaxes.  \citet{gizis} assembled almost 100 ground-based
parallaxes for SDs.  Brown-dwarf parallaxes are
at present obtained exclusively from the ground \citep{dahn}.

While nothing can replace this crucial ground-based work, at least until
a new generation of astrometry satellites is launched, it is in fact
possible to obtain many additional parallaxes of dim stars using
{\it Hipparcos} data.  Common proper-motion (CPM) companions of 
{\it Hipparcos} stars should have the same parallaxes as their 
primaries up to a fractional error equal to their separation in
radians, which is generally small compared to the measurement error
and, in any event, always less than 1\%.  

The idea of obtaining parallaxes in this way is not new.  In describing
the technique he had used to discover what was then (and remained for
four decades) the dimmest known star, vB 10, \citet{vB} recounted:
``In 1940 the author started at the prime focus of the 82-inch
reflector of the McDonald Observatory a systematic search
for faint companions to known PM stars in order to extend our knowledge
toward the lowest luminosity stars.''   The Yale Parallax Catalog
\citep{yale} lists the proper motions of CPM pairs together with
the parallax of the primary, clearly intending the same distance
estimate to be applied to the fainter star as well.  \citet{oppen01}
applied this technique to CPM companions of {\it Hipparcos} stars
to establish distances to some stars in their
8 pc sample.  However, no one has attempted to systematically search
for CPM companions of {\it Hipparcos} stars.  Indeed the Villanova White 
Dwarf web 
site\footnote{http://www.astronomy.villanova.edu/WDCatalog/index.html}
continues to list Yale parallaxes for CPM companions of stars that
now have {\it Hipparcos} parallaxes.

The primary problem in applying this technique
is establishing a physical association between
the two components of the binary.  This is not difficult for
very nearby stars, which usually have very large proper motions:
the chance that two unrelated stars lying within a few arcmin of each other
would have roughly similar proper motions of order $\mu\sim 1000\,\masyr$
is vanishingly small.  Hence, precise proper motions are not generally
required to establish a physical connection between these stars.
However, the projected density of stars as a function of proper motion
grows extremely rapidly toward lower proper motions, so much higher
precision is required to effectively reject spurious unrelated pairs.
Until recently, such high-precision proper-motion catalogs were not
available.

Several developments over the past 18 months have radically altered this
situation thereby permitting a much more aggressive search for CPM
companions of Hipparcos stars.  First, \citet{bright} and \citet{faint}
have published the revised New Luyten Two-Tenths (rNLTT) catalog,
which identifies virtually all {\it Hipparcos} counterparts of
NLTT \citep{luy,1st} stars, and which gives new more accurate 
$(\sigma_\mu\sim 5.5\,\masyr$) proper motions for the vast majority of 
NLTT stars in the 44\% of the sky covered by the intersection of the first 
Palomar Observatory Sky Survey (POSS I) and the Second Incremental Release of
2MASS \citep{2mass}.  Very importantly in the present context,
each rNLTT entry indicates whether the given star has a NLTT CPM companion
according to the NLTT Notes compiled by Luyten and, if so, whether that 
companion
is identified in rNLTT itself.  Moreover, for entries with CPM companions,
it gives the positional offset of the companion, both as given by
the NLTT Notes and as measured by rNLTT itself.  

Second, 2MASS has now been released for the whole sky.  Hence, by searching
2MASS at the CPM offsets as indicated by rNLTT, one can usually find the
CPM companions of {\it  Hipparcos} stars, at least those that satisfy the NLTT
proper motion threshold, $\mu\ge 180\,\masyr$.

Third, USNO-B1.0 \citep{usnob}, with its roughly $10^9$ proper motion 
measurements, each 
having typical errors of a few $\masyr$, has now been released.  
Because more than 99\% of its high proper-motion entries are spurious
\citep{gould03},
USNO-B cannot be used to search blindly for high proper motion stars.  However,
if one knows the approximate position and proper motion of a star (as one
does for CPM companions of {\it Hipparcos} stars) then the false background
stars become much more manageable.

Fourth, \citet{cg} have compiled a catalog of NLTT binaries for the 44\%
of the sky covered by the intersection of POSS I and the Second 2MASS 
Incremental Release.  For these areas, they have determined which
NLTT pairs that are specified in the NLTT Notes as CPM binaries are 
in fact physical pairs and also which pairs of NLTT stars are CPM
binaries despite the fact that they are not so designated in the Notes.
Hence, for these areas, virtually all the work required to assemble
a catalog of {\it Hipparcos} companions is already done.

In this paper, we make use of these various new data sources to compile
a catalog of CPM companions to {\it Hipparcos} stars.  The catalog is
restricted to {\it Hipparcos} stars that have accurate parallaxes and
are in the NLTT.  However, we also discuss
how it might be extended to companions of other {\it Hipparcos} stars
in future work.

\section{Catalog Construction
\label{sec:construct}}

\subsection{Philosophy
\label{sec:philosophy}}

Our aim is to construct a catalog with as many genuine CPM companions
as possible, while minimizing the number of false entries.  The quality
of our underlying sources varies dramatically over the sky, and so
we do not aim to construct a catalog that is ``complete'' or ``homogeneous''
in any sense.  When we have excellent proper-motion information, we
can be very confident that we have identified physical pairs.  As we will
describe, the available data are often far from adequate to achieve this
high level of confidence.  Hence, for many stars we must make a judgment
call, making use of photometric as well as astrometric data.  As this
photometry is generally photographic, it is sometimes not as reliable as one
would like.  Hence, any individual CPM pair in the catalog may
be spurious.  We expect that the primary use of the catalog will be
as a source of candidate dim stars with parallaxes, and that few will be
found to be spurious when checked by obtaining radial velocities and/or
CCD photometry.

	We exclude most pairs that are so close that they are not 
resolved in 2MASS.  These are so close that the proximity of their components
is usually a more reliable guide to their physical association 
than is the similarity
of their proper motions.  Hence, our proper-motion based approach brings
nothing new to the table.  Moreover, the great majority of these companions are
luminous stars (e.g., $M_V\sim 5$) that are already well represented in the 
{\it Hipparcos} catalog and hence are not of much interest in the present
context.  We make an exception to this rule only when the
companion is very dim and hence is a rare object from a parallax perspective.

	We count stars as being ``{\it Hipparcos}'' only if their parallax
measurements satisfy $\pi/\sigma_\pi >3$.  Otherwise, they do not have
$3\,\sigma$ parallax detections and therefore cannot provide significant
parallax information about their companion.  On the other hand, if such
stars are companions to other {\it Hipparcos} stars with good parallaxes,
they can gain parallax information.  In our search, we therefore treat
{\it Hipparcos} stars with less precise parallaxes as ``non-{\it Hipparcos}''.

	For completeness, we include CPM pairs of {\it Hipparcos} stars
that both have good parallaxes.  However, we consider these to be
of less (or at any rate, different) intrinsic interest and report them
in a separate table.  Of the 508 CPM pairs presented here, 84 are in
this category.

\subsection{Search Breakdown
\label{sec:breakdown}}

We divide our search for {\it Hipparcos} CPM companions into four
subcategories. First, pairs from the \citet{cg}
 catalog, which 
covers 44\% of the sky (248).  Second, other pairs from rNLTT for which
the primary (i.e., {\it Hipparcos} star) has a 2MASS identification in
rNLTT (16).  Third, other pairs from rNLTT for which the primary lacks a 
2MASS identification (88).  Fourth, pairs designated as CPM binaries by 
Luyten,
for which one star is in {\it Hipparcos} (and is in rNLTT) but for which
the companion is not in rNLTT (156).  As we will describe, these subcategories
require progressively more work.

\subsection{Catalog of Chanam\'e \& Gould (2003)
\label{sec:cg}}

   Chanam\'e \& Gould (2003, hereafter CG) have already done most
of the work to determine which Luyten CPM binaries are genuine and
have also found additional genuine CPM binaries among NLTT stars
that were not recognized as such by Luyten.  However, since the
objectives of CG and the present work are slightly different,
some additional work is required.  First, CG were interested
primarily in obtaining clean samples of disk and halo stars and so
excluded pairs that could not be reliably classified as one or the
other.  For example, those that straddled the disk/halo boundary
on a reduced proper motion (RPM) diagram were excluded.  They also
excluded all pairs for which one component was a WD.  Note
that MS pairs and SD pairs can be vetted by checking
to see whether they lie parallel to these respective sequences on an
RPM diagram (see their figs.\ 2 and 3), 
but that MS/WD pairs cannot be subjected to this test.
Thus, CG were not in a position to supply as severe vetting of
pairs with WD components as they were for other pairs and did not attempt
to do so.

Hence, our approach is to accept all pairs 
(having an {\it Hipparcos} component) regarded as genuine by CG
and then to review all candidates that they did not accept to see
if they should be regarded as genuine pairs.  This review makes use not
only of the information used by CG, but also of the {\it Hipparcos}
parallax.  For example, if a MS/WD pair is genuine, then when the WD
is placed at its companion's distance, it should lie close to the
WD sequence.

Finally, CG excluded all triples, including NLTT pairs for which
one NLTT component was resolved by the Tycho Double Star Catalog 
(TDSC, \citealt{tdsc}).
These do not appear in their catalog, even as candidates.  We make
no special effort to recover these as part of this subsearch: they
are recovered automatically as part of the other subsearches.

The proper-motion selection criterion of CG was to accept all
pairs with vector proper-motion differences satisfying 
$\Delta \mu \equiv |\Delta \vec\mu| < 20\,\masyr$, to further
accept those with $\Delta\mu < (112 - 51\log(\Delta\theta/''))\,\masyr$,
and to accept all pairs with separations $\Delta\theta<10''$.
In constructing these relatively severe criteria, CG took
advantage of the fact that rNLTT contains independent proper-motion
measurements for the  great majority of its stars and that the
{\it relative} proper-motion errors for these are quite small
(see \citealt{faint}).  

In the present context, however, it is important
to note that for some of CG's pairs, one component lacked an
independent proper-motion measurement.  For these, $\Delta \mu$ was determined
by taking the difference in the vector separations of the components
as recorded in the NLTT Notes and as measured by 2MASS and then dividing
this difference by the approximately 
45-year difference in epochs.  The errors from this
method are about twice as large as from direct measurements and,
moreover, have significant outliers, which probably originate
from transcription errors in the NLTT Notes.  This is a relatively
minor problem for this CG subsample: of the 248 binaries
that we eventually accept from this subcategory, only 39 lacked
independent proper-motion measurements.  All of these satisfied the
original $\Delta\mu$ criterion of CG.  However, in other subcategories,
notably the fourth, a much larger fraction of the binaries lack
independent proper-motion measurements.  We describe our procedure for
dealing with these below.

\subsection{Non-CG pairs in rNLTT with 2MASS data
\label{sec:noncg2m}}

A total of 16 such pairs are selected 
from two ultimate sources, Second Incremental 2MASS areas south
of POSS I (5), and pairs rejected by CG because they were part of triples
(11).  The first group is quite small because this area covers only 3\% of
sky.  All information is already available to classify these pairs.

\subsection{Non-CG pairs in rNLTT without 2MASS data
\label{sec:noncgnon2m}}

By construction, these pairs must have both components in one of three
position-and-proper-motion catalogs of bright stars,
{\it Hipparcos}, Tycho-2 \citep{t2}, or Starnet \citep{starnet},
which were the only avenues into rNLTT for stars in non-2MASS areas.
These stars almost all have very accurate proper
motions, and it is a simple matter at this point to search for them
in the 2MASS all-sky release.  Hence, their classification as genuine
or not is straightforward.  There are a total of 88 physical pairs, of which 
45 have both components in {\it Hipparcos}.  

Very few of these are of interest as dim stars with new parallaxes 
simply because they are so bright that if they had been of special
interest, they would have been observable directly with {\it Hipparcos}.

\subsection{Non-rNLTT CPM Companions of Hipparcos Stars
\label{sec:nonrnltt}}

This subcategory presents the greatest difficulties.  Given that,
even in cases for which the companion is not recovered by rNLTT,
the {\it Hipparcos} stars with NLTT companions are already marked
in rNLTT, it is straightforward to find the vast majority of these
in 2MASS simply by searching at the vector separation given in the
NLTT Notes (and recorded in rNLTT).  We then use the 2MASS position
to search for the companion in USNO-B.  When the companion is in
USNO-B, we obtain a second estimate of the proper motion in addition
to the one based on the difference between vector separations in
2MASS and the NLTT Notes (see \S~\ref{sec:cg}).  Whenever the
smaller of these two estimates violates the CG proper-motion
criterion (see \S~\ref{sec:cg}), we flag the binary.  We inspect
the RPM diagram of each binary and if the binary is not well
aligned with either the MS or SD tracks, we also flag it.  In particular,
all WDs are thus flagged.  Finally, we flag all binaries for which
one component is lacking 2MASS data.  Of the 184 initial candidates, a 
total of 82 are flagged, some several times.  We then review each of
the flagged cases individually, inspecting the binary on both
the RPM diagram, and the color-magnitude diagram (CMD), the latter
under the provisional assumption
that the {\it Hipparcos} parallax applies to both components.  Whenever
the binary is under question primarily because its proper-motion difference
is too high, we attempt to locate the companion in USNO-A
\citep{usnoa1,usnoa2}.  Comparison of this position with the 2MASS
position gives another estimate of the proper motion, which is
generally more reliable than either USNO-B or the 2MASS/NLTT-Notes
difference method \citep{faint,gould03}.  Unfortunately,  this method
usually fails for stars south of POSS I because they are usually not
in USNO-A.  See \citet{faint}.  In the end, we take all the available
evidence and make our best judgment as to whether the binary is
physical.  Of the 82 flagged binaries, 28 are rejected and another
10 are regarded as plausible but not fully convincing cases.  They
(along with one other pair from the CG subsample) are flagged as 
``somewhat uncertain'' in the catalog.

\section{Catalog Description
\label{sec:descript}}

The catalog is divided into two tables.  Table 1 lists the non-{\it Hipparcos}
CPM companions of {\it Hipparcos} stars, i.e., the stars with new 
{\it Hipparcos}-based parallaxes found by this work.  
Columns 1 through 9 describe the
companion: column 1 gives the NLTT number, columns 2 and 3 give the
R.A. and Dec (2000 epoch and equinox), columns 4 and 5 give the
east and north proper-motion components in $\masyr$.  Columns 6 and 7
give the $V$ magnitude and $V-J$ color, and column 8 gives the absolute
magnitude.  Column 9 is a 3 digit source code, which is described below.
Columns 10--18 are similar to columns 1--9, but for the {\it Hipparcos}
primary.  Columns 19 and 20 give the separation (in arcsec) and position angle
(north through east in deg) of the companion with respect to the 
{\it Hipparcos} star.  Columns
21--23 gives the {\it Hipparcos} number, parallax and parallax error
(in mas).  Column 24 is the adopted proper-motion difference and column
25 is a flag, ``1'' if the binary is ``somewhat uncertain'' and ``0''
otherwise.

The three-digit source code is an expanded version of the source code
used in rNLTT.  There, the three digits refer to the sources of the position,
proper motion, and $V$ photometry.  1 = Hipparcos, 2 = Tycho-2, 
3 = TDSC, 4 = Starnet,
5 = USNO/2MASS, 6 = NLTT, 7 = USNO-A (for position) or common proper motion 
companion (for proper motion).  More specifically, ``555'' means
2MASS based position, USNO-A based $V$ photometry, and USNO/2MASS based
proper motion.  In addition, we add 8 = USNO-B (for position and proper
motion).  A ``9'' in the position column means that there is no actual
detection in any catalog and the position is inferred from the separation
vector given in the NLTT Notes.  USNO-A photometry and NLTT photometry
are transformed to $V$ photometry using the prescriptions of \citet{faint}.

Table 2 lists the CPM pairs of {\it Hipparcos} stars, i.e., CPM pairs
composed of two {\it Hipparcos} stars, each with a parallax 
better than $3\,\sigma$.  
Column 1--6
give information on the brighter component.  Column 1 and 2 give the
{\it Hipparcos} and rNLTT numbers.  Column 3 and 4 give the parallax
and parallax error (in mas).  Columns 5 and 6 give the $V$ magnitude
and $V-J$ color.  Columns 7--12 give the same information for the
fainter companion.  Columns 13 and 14 give the separation and position
angle of the fainter component.  

These two tables are available at
\hfil\break\noindent
http://www.astronomy.ohio-state.edu/$\sim$gould/rNLTT\_binaries/new\_parallaxes.dat and
\hfil\break\noindent
http://www.astronomy.ohio-state.edu/$\sim$gould/rNLTT\_binaries/hip\_doubles.dat
respectively.
The Fortran format statements for the table records are respectively,
\hfil\break\noindent
(2(i5,2f10.5,2i6,3f7.2,i4,1x),2f6.1,i7,2f7.2,i5,i2) and
(2(i6,i6,2f7.2,2f7.2,1x),2f6.1).

\section{Dim Stars
\label{sec:Dim}}

\begin{figure}
\plotone{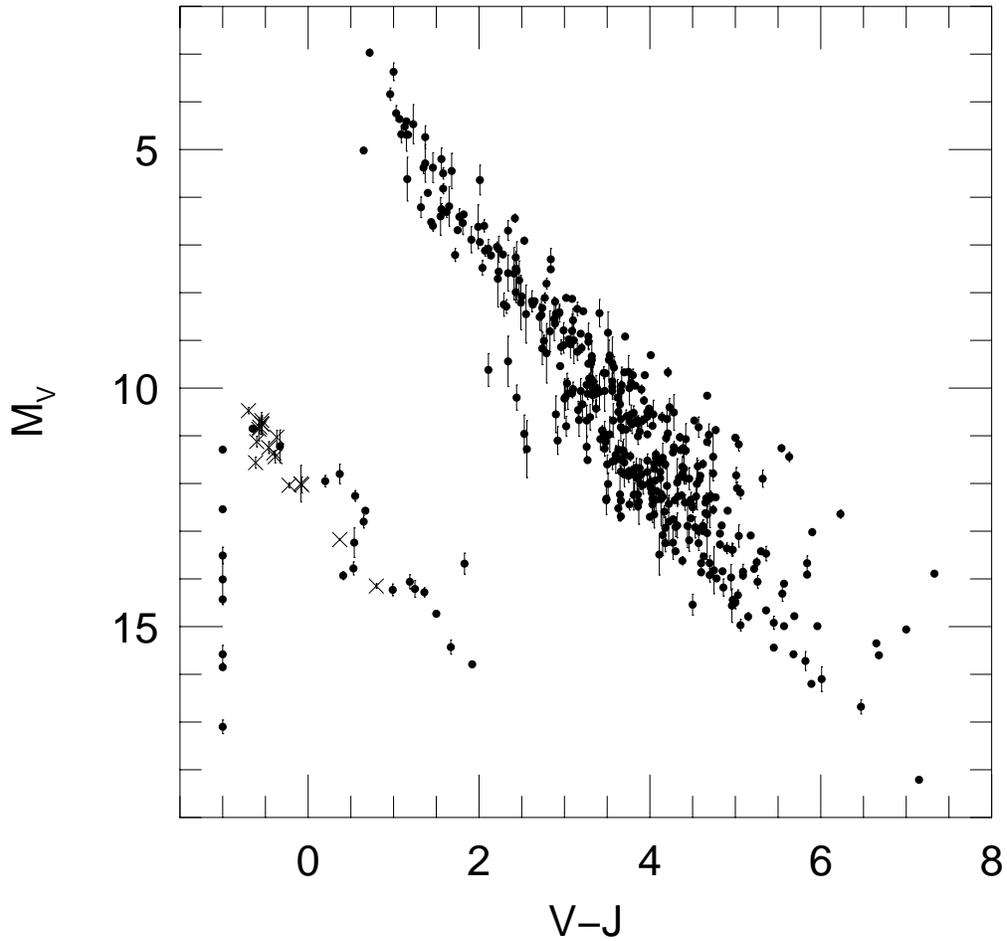}
\caption{\label{fig:cmd}
Color-magnitude diagram of 424 common proper motion (CPM) companions of
{\it Hipparcos} stars with good ($>3\,\sigma$) parallaxes.  Error
bars reflect only the parallax errors (and not the photometry
errors) and so reflect the precision {\it possible} if the present,
mostly photographic, photometry is replaced by CCD photometry.
Stars without $J$ photometry are displayed at $V-J=-1$.  In addition
all {\it Hipparcos} WDs with good parallaxes are shown as crosses.  
Note that these tend
to be significantly more luminous than the CPM WDs from the catalog.
}\end{figure}

\begin{figure}
\plotone{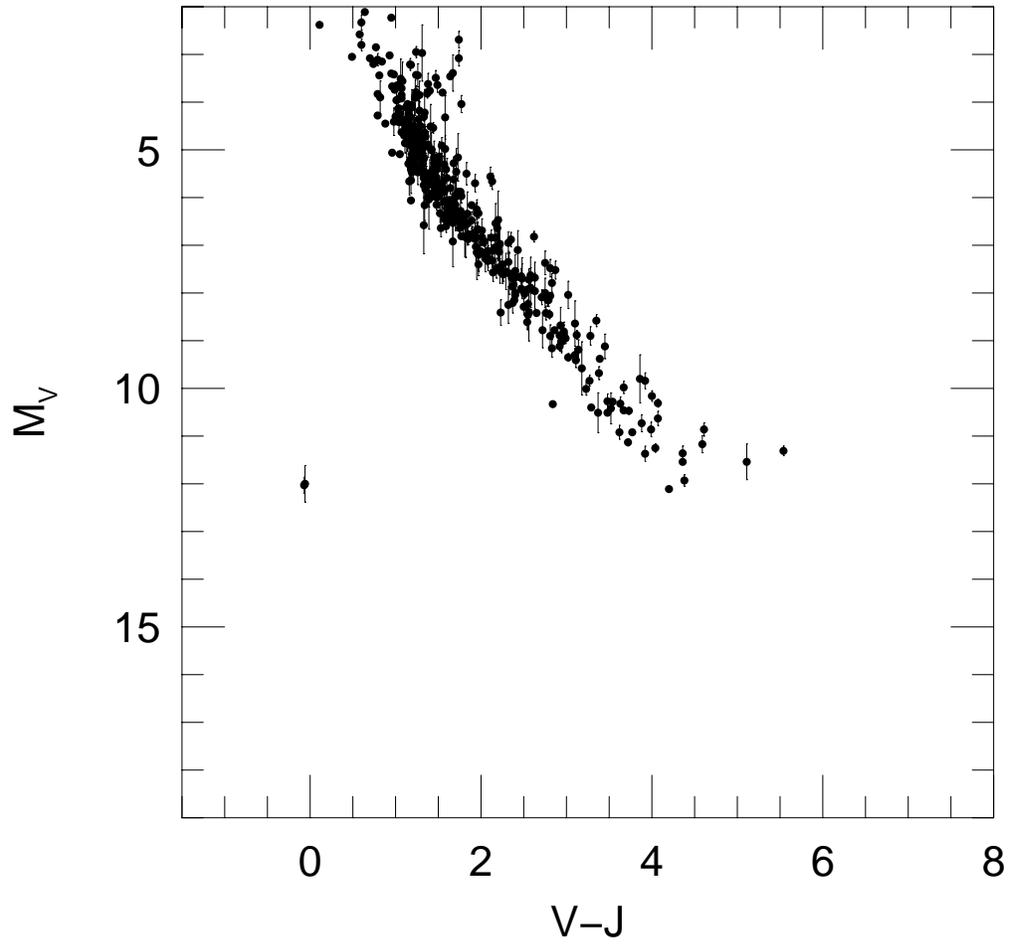}
\caption{\label{fig:cmd2}
Color-magnitude diagram of the {\it Hipparcos} primaries to the stars 
shown in Fig.~\ref{fig:cmd}.  Note that the CPM companions from 
Fig.~\ref{fig:cmd} tend to fill in the region at the lower right,
which is devoid of {\it Hipparcos} primaries.
}\end{figure}

Figure~\ref{fig:cmd} is a CMD of the CPM companions
of {\it Hipparcos} stars from Table 1.  The error bars reflect the
parallax errors only.  That is, they show the limits of precision for
the absolute magnitudes provided that one obtained good CCD photometry
for the stars.  The vast majority of the optical photometry is at present
photographic.  Stars without $J$ photometry are shown as $V-J=-1$.
These could be either M dwarfs or WDs.  Note the high concentration
of dim stars in the sample.  They are, for example, much dimmer than
their primaries, which are shown in Figure \ref{fig:cmd2}. 
Figure \ref{fig:cmd} also shows the
WDs from the {\it Hipparcos} catalog.  While {\it Hipparcos}
contains roughly the same number of WDs as our catalog, ours tend
to be dimmer.

A striking feature of the CMD is its breadth, roughly 3 mag.  This is
much broader than the CMD of \citet{monet92}, which is likely to be
due in significant part to errors in the photographic photometry.
However, the underlying proper-motion limited sample is likely to
have a broader range of metallicities than typical samples, and this
may also contribute to the breadth of the CMD.  It would be straightforward
to obtain CCD photometry for the entire sample and more than half could
be done in a single few-night run on a 1m telescope.  Perfect photometric
conditions would not be required to achieve dramatic improvements over
the present optical photometry.

\begin{figure}
\plotone{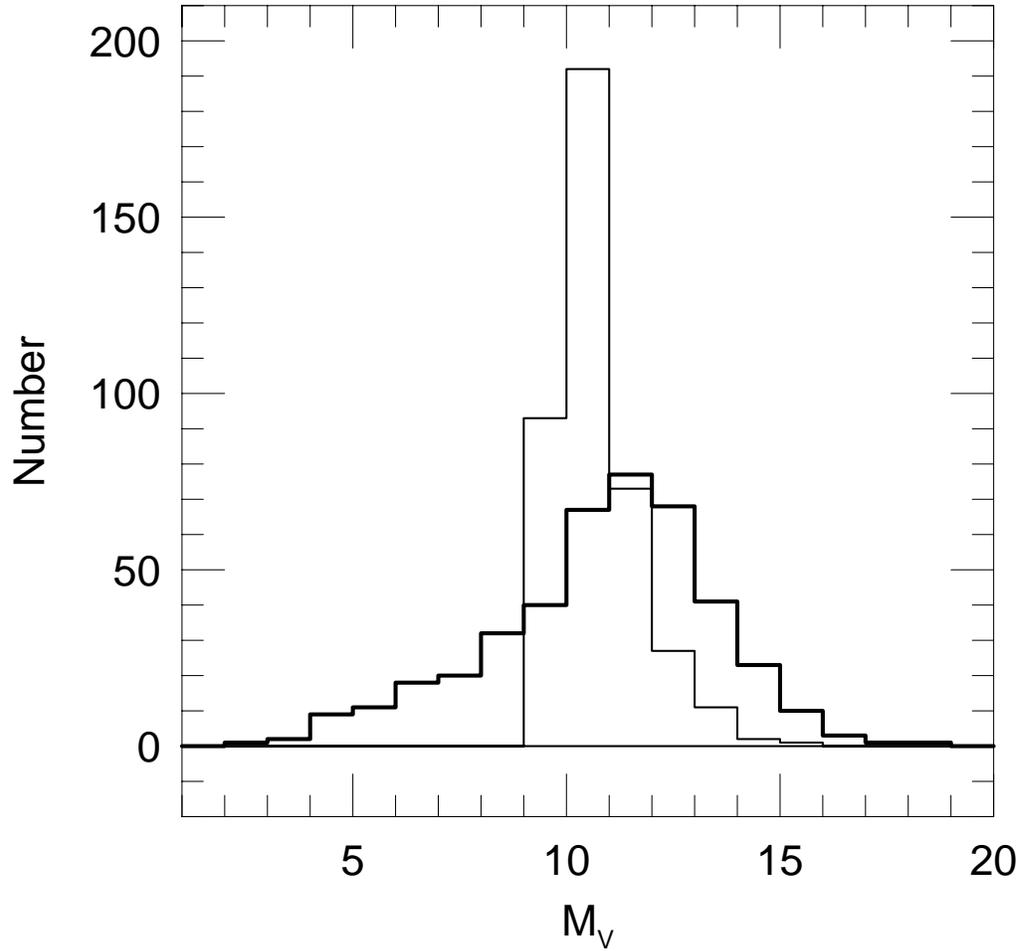}
\caption{\label{fig:dimhist}
Distribution of absolute magnitudes of the 424 stars in the catalog of CPM 
companions to {\it Hipparcos} stars ({\it bold histogram}), compared to
that of the 400 dimmest stars in {\it Hipparcos} itself ({\it solid
histogram}).  Note that the CPM catalog contains many more extremely
dim stars.
}\end{figure}

In Figure~\ref{fig:dimhist}, we compare histograms of the stars in Table
1 and of the 400 dimmest stars in the {\it Hipparcos} catalog (restricted
to entries with $\pi/\sigma_\pi>3$).  While both catalogs contain many
moderately dim stars, our catalog contains substantially more extremely 
dim stars than {\it Hipparcos}.

What was previously known about these dim CPM companions of 
{\it Hipparcos} stars?  We address this question in Tables 3 and
4.  Table 3 lists the 20 CPM WD companions shown in Figure~\ref{fig:cmd}
and indicates whether each was previously identified as being a WD
and, if so, whether it had a measured parallax.  The classifications
are taken from the Villanova White Dwarf web site
except for NLTT 49859, which comes from 
SIMBAD\footnote{http://simbad.u-strasbg.fr/Simbad}.  The 
previous parallaxes come most directly from Villanova, but most
come ultimately from \citet{yale}, with the one exception (NLTT 15768)
coming from \citet{bergeron01}.  The identifiers come mostly from 
SIMBAD, except for NLTT 26462 and 29967, which come from Villanova.
Column 1 gives the NLTT number.  
Columns 2 and 3 give the RA and Dec (2000 epoch and equinox).  
Columns 4--6 give the $V$, $V-J$, and $M_V$ magnitude, color,
and absolute magnitude.
Columns 7 and 8 give the separation and position angle relative
to the {\it Hipparcos} star, whose {\it Hipparcos} number, parallax,
and parallax error are given in columns 9--11.  Columns 12 and 13 give 
the previously tabulated parallax and error for the WD if any were found.
Column 14 gives one of the identifiers (if any were found)
and Column 15 gives the spectral type.  Of the 20 WDs,
five (NLTT 12412, 38926, 42785, 47097, 49859) are not listed at Villanova
as WDs.  Two of these are listed but
not classified by SIMBAD, while NLTT 12412 and 38926 are not listed 
and NLTT 49859
is classified as an M dwarf.
Of the remaining 15 WDs, seven have previous parallaxes of which only two
are of comparable quality to the {\it Hipparcos} parallaxes assigned in
the present work.

With one exception, the stars in Table 3 are classified as WDs based
either on their position on the CMD or because they were so classified
in the Villanova White Dwarf web site.  The exception is NLTT 38926.
We classify this as a WD in spite of the fact that we have no $J$
data and the only available reference in the literature (Luyten) classified
it as an ``m'' star.  We do so because of its dim absolute magnitude, 
$M_V=13.5$,
and the fact that it is significantly bluer than 4 random neighboring
field stars in $B-R$ and $R-I$, based on Digitized Sky 
Survey\footnote{http://archive.eso.org/dss/dss} images.
However, 
we record Luyten's contrary view in the
spectral classification column with the denotation ``L:m''. 

Table 4 lists the 29 stars with $M_V\geq 14$ from Table 1, except that
the WDs listed in Table 3 are excluded.  The columns in Table 4 are the
same as those in Table 3.  Five of these 29 stars have previously
recorded parallaxes, of which three (NLTT 923, 26247, 47621)
were determined by \citet{oppen01} based on their being CPM companions
to {\it Hipparcos} stars, and the other two (NLTT 18218, 42494) were
determined by \citet{yale} based on their being CPM companions to brighter 
stars with parallaxes.  Only 12 of the 29 have spectral types listed in
SIMBAD, and many are not listed at all (and hence have no identifier
in Table 4).
Three of the stars (NLTT 28864, 40719, 41096)
have no $J$ data, and moreover have no classification in the literature.
For these, we list the Luyten classification based on photographic
colors (e.g., ``L:f'', meaning ``F star'').  If these classifications
are correct, then
two of these stars (NLTT 28864, 41096) are WDs.  We confirm the existence
of NLTT 41096 based on Digitized Sky Survey images, but the other
two stars are too close to their companions to be seen.  The identifier
and spectral class of NLTT 8870 are assigned question marks because
the catalogued star with this identifier has roughly the right characteristics
but the wrong position in SIMBAD.

\section{Hipparcos Doubles
\label{sec:doubles}}

\begin{figure}
\plotone{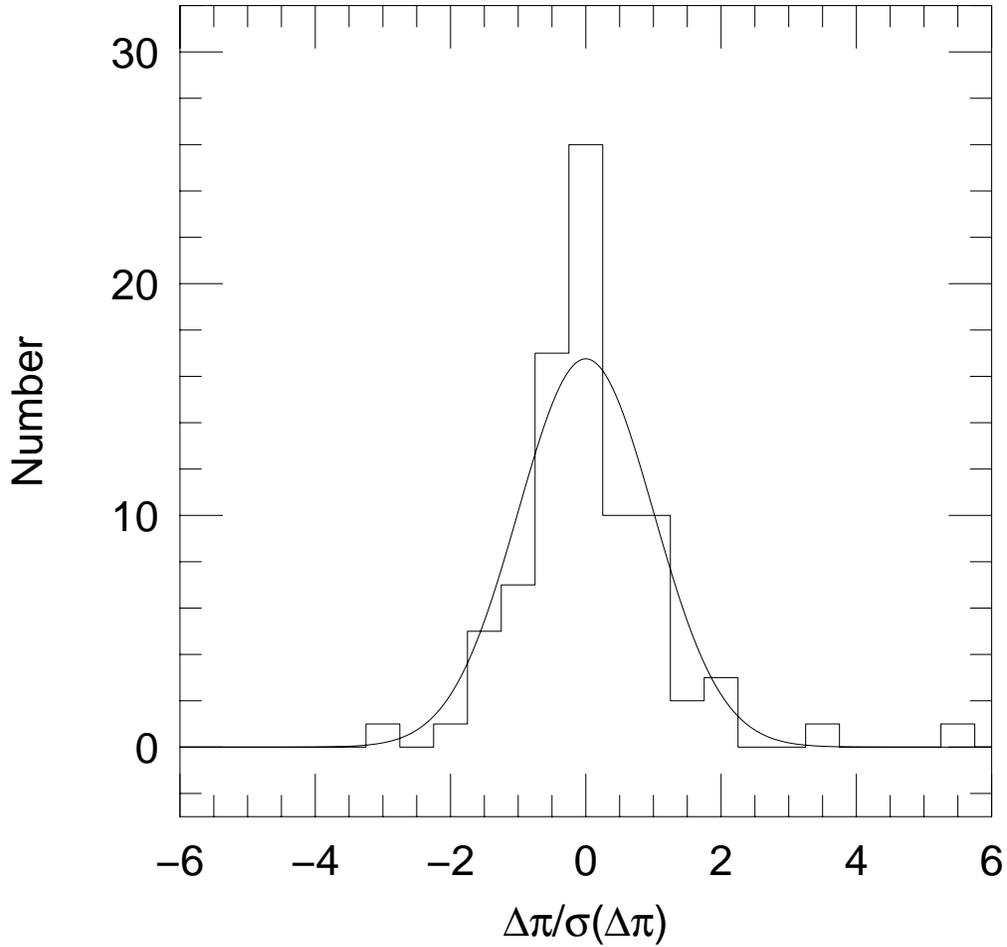}
\caption{\label{fig:errdist}
Differences between {\it Hipparcos} parallaxes divided by the root-sum
square of the errors of the two components
for 84 binaries with both components in {\it Hipparcos}.  Histogram
shows the actual distribution while the curve shows the expectation
based on Gaussian statistics.  There is rough agreement, but the
actual distribution contains 4 pairs with exactly zero difference and
three outliers at $2.8, 3.6,$ and $5.3\,\sigma$, neither of which
can be explained on the basis of Gaussian statistics.
}\end{figure}

The pairs in Table 2 are of modest interest.  For a few, one can
obtain a significantly better parallax for one star by using
the parallax of the other.  However, one important application
of this table is to test the accuracy of the error bars listed
in the {\it Hipparcos} catalog.  Figure~\ref{fig:errdist} is a histogram
of the differences (brighter minus fainter) in the parallaxes
of the binary components divided by the root sum square of their
reported errors.  The curve shows the distribution expected for
Gaussian errors.

Figure~\ref{fig:errdist} has three notable features.  First, the
curve is in overall rough agreement with the histogram, indicating
that, on average, the errors are properly estimated.  Second, there
is a spike at zero difference.  Third, there are two severe
outliers at $+3.6\,\sigma$ and $+5.3\,\sigma$, and a third mild
outlier at $-2.8\,\sigma$ (respectively Hip 82817/09, 90355/65,
64444/3).

The spike is principally due to four binaries with exactly zero
difference in parallax (Hip 13714/6, 17749/50, 86961/3, 97099/6).
Indeed the two components of these four pairs have exactly the
same reported errors as well.  While the {\it Hipparcos} errors
are known to be correlated at close separations (e.g., \citealt{ng}),
the degree of correlation observed in these four pairs
would appear a bit extreme.  It is
not clear what singles out these four. While they all have
separations $\Delta\theta\sim 20''$, and so are closer than
typical binaries in Table 2, there are 14 other binaries in
this table with $\Delta\theta< 20''$, and these appear to
have normally distributed errors.

Each of the three outliers has components with almost identical
proper motions and so can be taken to be a genuine physical pair
with virtual certainty.  None of the six components of
these pairs is resolved in TDSC as two distinct stars.

\section{Future Possibilities
\label{sec:future}}

The technique developed here could be applied to other data sets
to obtain parallaxes for additional dim stars.  Here we briefly
outline the potential possibilities and pitfalls of this approach.

The key characteristic of the {\it Hipparcos}-CPM binaries reported here 
that made them relatively easy to find is that someone (mostly Luyten, 
but in a few cases CG) had already tabulated them as probable binaries
and had recorded separations and position angles.  It was then
relatively straightforward to find the CPM companions in various catalogs 
and to use the data so obtained to judge the genuineness of the candidates.

Hence, one would be well advised to apply the same approach to
another catalog of wide binaries, the most obvious choice being
the Luyten Double Star (LDS) Catalog \citep{lds}.  This is comprised primarily
of candidate wide binaries found by Luyten in his search for
high proper-motion stars ($\mu\geq 180\,\masyr$), but including even those
that did not meet this threshold.  
The only additional required step (relative to
the work reported in this paper) would be to match LDS primaries
to {\it Hipparcos} stars.  This was unnecessary for the NLTT binaries
because their {\it Hipparcos} counterparts had already been identified
by rNLTT.  The drawback of LDS is that it will most likely yield
a lower fraction of dim stars: for dim stars to be recovered in
NLTT or LDS, they must be relatively close to satisfy the
$V\sim 19$ magnitude limit of these catalogs.  They then typically
have high proper motions, and so would tend to be in NLTT as well as LDS.

Another option would be to search in USNO-B for CPM companions of
{\it Hipparcos} stars.  While USNO-B cannot be used for a blind
search because most of its high proper-motion entries are spurious
\citep{gould03}, if the search is restricted to the narrow range
of proper motions compatible with the known proper motion of
an {\it Hipparcos} star, the number  of spurious entries can
be drastically reduced.  Nevertheless, CG found a similar
search for NLTT CPM companions in USNO-B to be quite
laborious because of the still large number of spurious
entries.  Moreover, the return of real binaries (not already
in NLTT) was relatively low.  See their figure 13.  On the
plus side,  however, most of those found were fainter than
the NLTT companions, and so (at fixed distance) would also
be dimmer.  The false detections from  such a USNO-B search
could be drastically reduced by demanding detection of the
candidates in 2MASS.  Unfortunately, this would have the effect
of eliminating most WDs, although it would preserve most red
dwarfs.

\acknowledgments 
We thank Samir Salim for making several very valuable comments
on the manuscript.
We thank D. Monet and the USNO-B team for providing us with a copy of
the USNO-B1.0 catalog.  This publication has made use of catalogs from
the Astronomical Data Center at NASA Goddard Space Flight Center,
VizieR and SIMBAD databases operated at CDS, Strasbourg, France, and
data products from the Two Micron All Sky Survey, which is a joint
project of the University of Massachusetts and the IPAC/Caltech,
funded by NASA and the NSF.  The ESO Online Digitized Sky Survey is
based on photographic data obtained using The UK Schmidt
This work was supported by grant AST 02-01266 from the NSF 
and by JPL contract 1226901.


\clearpage
\renewcommand{\arraystretch}{.6}
\begin{landscape}
\begin{deluxetable}{rrrrrrrrrrrrrll}
\tablewidth{0pc}
\tablecaption{White Dwarf Companions to Hipparcos Stars}
\tablehead{
 \colhead{NLTT}  
& \colhead{RA}  
& \colhead{Dec}  
& \colhead{$V$}
& \colhead{$V$-$J$}
& \colhead{$M_V$}
& \colhead{$\Delta\theta$}
& \colhead{p.a.}
& \colhead{Hip.~\#}  
& \colhead{$\pi$}
& \colhead{$\sigma(\pi)$}
& \colhead{$\pi_{\rm old}$}
& \colhead{$\sigma(\pi)$}
& \colhead{Name}
& \colhead{Type}
\\
\colhead{} & 
\colhead{(2000)} & 
\colhead{(2000)} & 
\colhead{} & 
\colhead{} & 
\colhead{} & 
\colhead{$"$} & 
\colhead{deg} & 
\colhead{} & 
\colhead{mas} & 
\colhead{mas} & 
\colhead{mas} & 
\colhead{mas} & 
\colhead{} & 
\colhead{} 
}
\startdata

 1762 &   8.26164 &  44.73697 & 16.33 & -0.33 & 11.22 &  28.8 &  30.7 &   2600 &  9.52 &  1.63 &      &      & EGGR 306   & DA4\\
12412 &  60.49868 & -34.46993 & 17.82 &  0.99 & 14.23 &  64.1 & 313.0 &  18824 & 19.15 &  1.13 &      &      &            &    \\
13599 &  69.18708 &  27.16430 & 15.99 &  1.50 & 14.73 & 123.9 & 338.9 &  21482 & 56.02 &  1.21 & 60.2 &  2.9 & G 39-27    & DA \\
15768 &  88.47462 &  12.40402 & 15.60 &  0.20 & 11.95 &  90.0 & 232.0 &  27878 & 18.64 &  1.05 & 15.1 &  5.0 & EGGR 44    & DC5\\
16355 &  94.05943 & -59.20764 & 13.65 & -0.65 & 10.85 &  40.4 & 301.1 &  29788 & 27.50 &  0.50 & 42.0 & 15.0 & LTT 2511   & DB4\\
18414 & 116.41036 & -33.93111 & 16.70 &  1.92 & 15.79 & 870.9 &   2.8 &  37853 & 65.79 &  0.56 & 57.5 &  3.1 & GJ 288B    & DC9\\
26462 & 167.37547 & -26.01849 & 16.79 &  0.41 & 13.93 & 100.2 & 180.0 &  54530 & 26.76 &  1.09 &      &      & LP 849-059 & DC \\
29967 & 183.15007 &  -6.37170 & 17.26 &  1.19 & 14.06 & 202.8 & 102.4 &  59519 & 22.94 &  1.63 &      &      & LP 674-029 & DC \\
38926 & 224.46685 &  29.88101 & 17.66 & -9.00 & 13.51 &  27.5 & 314.0 &  73224 & 14.79 &  1.24 &      &      &            & L:m\\
41169 & 236.87544 & -37.91897 & 13.46 & -9.00 & 12.54 &  15.0 & 130.0 &  77358 & 65.60 &  0.77 & 74.6 & 10.1 & GJ 599B    & DA7\\
42785 & 246.57123 &   2.18123 & 17.42 &  1.83 & 13.68 &   9.3 & 319.5 &  80522 & 17.86 &  1.91 &      &      & LHS 3195   &    \\
47097 & 282.17372 &  68.87733 & 17.18 &  0.65 & 12.80 &  33.9 & 273.4 &  92306 & 13.33 &  0.85 &      &      & LDS 2421B  &    \\
49859 & 311.52328 &  33.97068 & 14.29 &  0.67 & 12.57 &  87.9 & 271.2 & 102488 & 45.26 &  0.53 &      &      & GJ 9707C   & MV4\\
50189 & 314.19911 &  -4.84440 & 16.67 &  1.67 & 15.43 &  14.5 & 309.8 & 103393 & 56.56 &  4.03 & 64.6 &  5.1 & vB 11      & DC9\\
51482 & 323.06767 &   0.25407 & 15.27 &  0.37 & 11.80 & 133.0 &  29.0 & 106335 & 20.26 &  2.00 &      &      & PHL 28     & DB \\
53526 & 334.86906 &  21.37246 & 17.69 &  1.25 & 14.21 &  83.3 & 282.3 & 110218 & 20.12 &  1.71 &      &      & LP 460-3   & DC \\
55288 & 343.95613 &  -7.83403 & 16.99 &  1.36 & 14.28 &  41.8 & 190.4 & 113231 & 28.72 &  1.30 & 36.7 &  5.3 & G 156-64   & DA \\
55300 & 343.98225 &   5.75618 & 15.71 &  0.53 & 13.78 &  17.1 & 285.2 & 113244 & 41.15 &  2.65 &      &      & GJ 4305B   & DB \\
55701 & 345.65880 &  76.50108 & 16.35 &  0.55 & 12.26 &  13.2 & 137.6 & 113786 & 15.22 &  0.81 &      &      & EGGR 454   & DC6\\
57958 & 356.79796 & -26.39293 & 16.83 &  0.54 & 13.24 &  13.2 & 139.2 & 117308 & 19.11 &  2.93 &      &      & G275-102   & DB \\

\enddata
\end{deluxetable}
\end{landscape}


\clearpage

\renewcommand{\arraystretch}{.6}
\begin{landscape}
\begin{deluxetable}{rrrrrrrrrrrrrll}
\tablewidth{0pc}
\tablecaption{$M_V>14$ Companions to Hipparcos Stars}
\tablehead{
 \colhead{NLTT}  
& \colhead{RA}  
& \colhead{Dec}  
& \colhead{$V$}
& \colhead{$V$-$J$}
& \colhead{$M_V$}
& \colhead{$\Delta\theta$}
& \colhead{p.a.}
& \colhead{Hip.~\#}  
& \colhead{$\pi$}
& \colhead{$\sigma(\pi)$}
& \colhead{$\pi_{\rm old}$}
& \colhead{$\sigma(\pi)$}
& \colhead{Name}
& \colhead{Type}
\\
\colhead{} & 
\colhead{(2000)} & 
\colhead{(2000)} & 
\colhead{} & 
\colhead{} & 
\colhead{} & 
\colhead{$"$} & 
\colhead{deg} & 
\colhead{} & 
\colhead{mas} & 
\colhead{mas} & 
\colhead{mas} & 
\colhead{mas} & 
\colhead{} & 
\colhead{} 
}
\startdata

  638 &   3.42971 &  80.66376 & 16.38 &  5.45 & 14.92 &  12.8 & 126.1 &   1092 &  51.12 &   3.31 &        &        & G 242-48B  & M5    \\
  923 &   4.60755 &  44.02723 & 12.75 &  5.96 & 14.99 &  35.1 &  64.0 &   1475 & 280.27 &   1.05 & 280.27 &   1.05 & HD 1326B   & M3.5  \\
 2859 &  12.89642 & -22.90851 & 16.35 &  5.45 & 15.44 &  16.8 &  71.4 &   4022 &  65.81 &   1.44 &        &        & LDS 1082B  &       \\
 8870 &  41.42178 &  44.95083 & 16.19 &  5.03 & 14.34 &  18.1 &  65.6 &  12886 &  42.59 &   2.21 &        &        & GJ 3179B ? & M5 ?  \\
 8960 &  42.42210 &  75.02788 & 17.24 &  5.55 & 14.31 &  28.5 &  57.6 &  13179 &  25.98 &   2.00 &        &        & LDS 1559B  &       \\
10761 &  50.63969 & -30.19058 & 17.20 &  5.15 & 14.79 &  24.9 & 253.1 &  15725 &  32.99 &   1.29 &        &        &            &       \\
15867 &  89.57157 &  -4.63374 & 16.49 &  5.36 & 14.66 &  89.3 & 313.1 &  28267 &  43.10 &   0.93 &        &        & BD-04 1310 &       \\
18218 & 114.90280 &  33.46461 & 18.83 &  6.01 & 16.10 &  13.0 &  57.4 &  37312 &  28.48 &   3.59 &  26.40 &   2.50 & GJ 3458B   &       \\
19472 & 126.21852 &  -3.68378 & 17.16 &  5.57 & 14.99 & 356.4 &  47.1 &  41211 &  36.75 &   0.87 &        &        &            &       \\
22015 & 143.20107 &  26.99552 & 16.03 &  5.69 & 14.78 &  65.2 &  67.3 &  46843 &  56.35 &   0.89 &        &        & GJ 354.1B  & M5.5  \\
24550 & 157.87822 &  57.08834 & 15.31 &  5.57 & 14.10 & 141.9 & 225.7 &  51547 &  57.18 &   1.31 &        &        & LDS 2314B  & M4.5  \\
26247 & 166.37876 &  43.52164 & 14.62 &  5.89 & 16.20 &  31.4 & 126.1 &  54211 & 206.94 &   1.19 & 206.94 &   1.19 & GJ 412B    & M6    \\
28453 & 176.39744 & -20.35138 & 18.20 &  6.47 & 16.68 &  15.4 &  60.8 &  57361 &  49.63 &   3.55 &        &        & LP 793-34  & late M\\
28513 & 176.63621 & -40.49657 & 15.43 &  6.68 & 15.60 &  22.9 &  53.5 &  57443 & 108.23 &   0.70 &        &        & vB 5       & M4    \\
28864 & 178.31940 &  -7.37521 & 18.67 & -9.00 & 17.10 &   9.0 & 110.0 &  57959 &  48.63 &   3.34 &        &        & GJ 452B    & L:f   \\
29580 & 181.44429 & -18.82557 & 16.20 &  4.97 & 14.44 & 187.7 & 342.3 &  59000 &  44.41 &   1.51 &        &        &            &       \\
29661 & 181.84996 &  13.03719 & 18.06 &  5.06 & 14.97 &  66.9 & 283.3 &  59126 &  24.08 &   1.38 &        &        &            &       \\
29896 & 182.85170 &  43.55919 & 18.61 &  4.50 & 14.54 &  10.5 & 228.7 &  59428 &  15.35 &   1.64 &        &        &            &       \\
32366 & 194.12921 &  21.00776 & 19.04 &  4.96 & 14.56 &  35.2 & 244.0 &  63168 &  12.74 &   2.24 &        &        & LDS 2891C  &       \\
34218 & 202.08720 &  30.05528 & 19.14 &  5.82 & 15.72 &  52.6 &  51.4 &  65706 &  20.74 &   1.97 &        &        & LDS 1390B  &       \\
35919 & 209.84001 &  25.23418 & 17.26 &  4.86 & 14.18 &  20.3 & 177.8 &  68342 &  24.23 &   2.09 &        &        & LHS 2838   & M     \\
40719 & 234.21327 &  37.58035 & 19.28 & -9.00 & 15.58 &  10.0 &  91.0 &  76452 &  18.21 &   1.65 &        &        &            & L:+m  \\
41096 & 235.54552 &  72.79309 & 18.06 & -9.00 & 14.01 &  15.7 & 358.0 &  76902 &  15.48 &   4.29 &        &        & LDS 1828B  & L:a-f \\
42209 & 242.42474 &  65.82235 & 18.96 &  5.00 & 14.50 &  29.0 & 147.2 &  79180 &  12.82 &   0.77 &        &        & LDS 2369B  &       \\
42494 & 245.01351 & -37.53030 & 14.99 &  6.65 & 15.35 &   5.3 & 227.2 &  80018 & 118.03 &   2.28 & 130.00 &   9.70 & GJ 618B    & M5.5  \\
47621 & 289.24002 &   5.15044 & 17.05 &  7.15 & 18.21 &  74.0 & 150.0 &  94761 & 170.26 &   1.37 & 170.26 &   1.37 & vB 10      & M8V   \\
49961 & 312.30737 &  32.28100 & 17.45 &  5.68 & 15.58 &  34.4 & 246.0 & 102766 &  42.23 &   0.98 &        &        & LDS 2931B  &       \\
53379 & 334.01075 &  54.66652 & 16.72 &  7.00 & 15.06 &  76.8 & 107.4 & 109926 &  46.62 &   0.67 &        &        & GJ 4269B   & M4    \\
57088 & 352.71868 &  27.77598 & 17.65 &  5.26 & 14.06 &  26.7 & 250.5 & 116052 &  19.13 &   1.21 &        &        &            &       \\
\enddata
\end{deluxetable}
\end{landscape}

\clearpage

\clearpage

\begin{thebibliography}{}

\bibitem[Bergeron, Leggett \& Ruiz(2001)]{bergeron01} 
Bergeron, P., Leggett, S. K., \&  Ruiz, M.T.\ 2001, \apjs, 133, 413.

\bibitem[Chanam\'e \& Gould(2003)]{cg} Chanam\'e, J. \& Gould, A.\
2003, \apj, submitted (astroph/0307284)

\bibitem[Dahn et al.(2002)]{dahn} Dahn, C.C., et al.\ 2002, \aj, 124, 1170

\bibitem[ESA(1997)]{hip} European Space Agency (ESA). 1997, The Hipparcos and
Tycho Catalogues (SP-1200; Noordwijk: ESA)

\bibitem[Fabricius et al.(2002)]{tdsc} Fabricius, C.,
H{\o}g, E., Makarov, V.~V., Mason, B.~D., Wycoff, G.~L., \& Urban,
S.~E.\ 2002, \aap, 384, 180.

\bibitem[Gizis(1997)]{gizis} Gizis, J.E.\ 1997, \aj,  113, 806

\bibitem[Gould(2003)]{gould03} Gould, A.\ 2003,  \aj, 126, 472

\bibitem[Gould \& Salim(2003)]{bright} Gould, A.\ \& Salim, S.\ 2003,
  \apj, 582, 1001


\bibitem[H{\o}g et al.(2000)]{t2} H{\o}g, E.~et al.\ 2000, \aap, 355, L27.


\bibitem[Jao et al.(2003)]{jao} Jao, W.C, Henry, T.J.,
 Subasavage, J.P., Bean, J.L.,  Costa, E., Ianna, P.A., \&
 M\'endez, R.A.\ 2003, \apj 125, 332

\bibitem[Luyten(1979, 1980)]{luy} Luyten, W.J.\ 1979, 1980, New Luyten
Catalogue of Stars with Proper Motions Larger than Two Tenths of an Arcsecond
(Minneapolis: University of Minnesota Press)

\bibitem[Luyten(1940-87)]{lds}Luyten, W.J.\ 1940-87, The LDS Catalogue:
Double Stars with Common Proper Motion 
(Minneapolis: University of Minnesota Press)

\bibitem[Luyten \& Hughes(1980)]{1st} Luyten, W.J. \& Hughes, H.S.\ 1980,
Proper Motion Survey wit h the Forty-Eight Inch Schmidt Telescope. LV. First
Supplement to the NLTT Catalogue (Minneapolis: University of Minnesota Press)

\bibitem[Monet(1996)]{usnoa1} Monet, D.G.\ 1996, American Astronomical
Society Meeting, 188, 5404.

\bibitem[Monet(1998)]{usnoa2} Monet, D.~G.\ 1998, American Astronomical
Society Meeting, 193, 112003

\bibitem[Monet et al.(1992)]{monet92} Monet, D.G., Dahn, C.C.,
 Vrba, F.J., Harris, H.C., Pier, J.R., Luginbuhl, C.B., \&
 Ables, H.D. 1992, \aj, 103, 638

\bibitem[Monet et al.(2003)]{usnob} Monet, D.G.\ 2003, \aj, 125, 984

\bibitem[Narayanan \& Gould(1999)]{ng}  Narayanan, V.K. \& Gould, A.\
1999, \apj, 523, 328

\bibitem[Oppenheimer et al.(2001)]{oppen01}Oppenheimer, B. R., Golimowski, 
D.A., Kulkarni, S.R., Matthews, K., Nakajima, T., Creech-Eakman, M., \&
 Durrance, S.T.\ 2001, \aj, 121, 2189

\bibitem[R{\" o}ser(1996)]{starnet} R{\" o}ser, S.\ 1996, IAU Symp.~172:
Dynamics, Ephemerides, and Astrometry of the Solar System, 172, 481.



\bibitem[Salim \& Gould(2003)]{faint} Salim, S.~\& Gould, A.\ 2003, \apj,
582, 1011

\bibitem[Skrutskie et al.(1997)]{2mass} Skrutskie, M.~F.~et al.\ 1997, in The
Impact of Large-Scale Near-IR Sky Survey, ed. F. Garzon et al (Kluwer:
Dordrecht), p.\ 187

\bibitem[van Altena, Lee, \& Hoffleit(1995)]{yale} van Altena, W.F.,
Lee, J.T., \& Hoffleit, E.D.\ 1995,
General Catalog of Trigonometric Parallaxes, Fourth Edition (Yale:
New Haven)

\bibitem[van Biesbroeck(1944)]{vB} van Biesbroeck, G.\ 1944, \aj, 51, 61

\end{thebibliography}
\end{document}